\documentclass[12pt]{article}

\usepackage{scicite}

\usepackage{times}

\usepackage{graphicx}
\usepackage{amsmath,amssymb}
\usepackage{bm}
\usepackage{siunitx}
\usepackage{booktabs}
\usepackage{multirow}
\usepackage{url}
\usepackage{gensymb}
\newcounter{lastnote}

\newenvironment{sciabstract}{%
\begin{quote} \bf}
{\end{quote}}

\vspace{-6.0cm}
\def\scititle{Electric field controlled spin transport in a topological insulator interfaced with a ferroelectric antiferromagnet}
\title{\scititle}

\author{
Yogesh Kumar,$^{1}$ Pushpendra Gupta,$^{2}$ Xinyan Li,$^{3}$ Richa Mudgal,$^{3,4}$\\ Ashish Omar,$^{2}$ Ryan Chen,$^{2}$
Mito Funatsu,$^{2}$ \\ Maya Ramesh,$^{5}$ Nicholas Reiterer,$^{2}$
Yuanqi Lyu,$^{6}$  Yiping Zeng$^{2}$\\ Darrell G. Schlom,$^{5,7,8}$
Alessandra Lanzara,$^{1,6}$ Robert J. Birgeneau,$^{1,6}$\\ James G. Analytis,$^{6}$
Ramamoorthy Ramesh,$^{1,2,6}$ Sajid Husain,$^{2,6\ast}$\\[6pt]
\normalsize{$^{1}$Materials Science Division, Lawrence Berkeley National Laboratory, Berkeley, CA,  USA.}\\
\normalsize{$^{2}$Department of Materials Science and Engineering, University of California, Berkeley, CA , USA.}\\
\normalsize{$^{3}$Rice Advanced Materials Institute, Rice University, Houston, TX , USA.}\\
\normalsize{$^{4}$Department of Materials Science and NanoEngineering, Rice University, Houston, TX, USA.}\\
\normalsize{$^{5}$Department of Materials Science and Engineering, Cornell University, Ithaca, NY, USA.}\\
\normalsize{$^{6}$Department of Physics, University of California, Berkeley, CA, USA.}\\
\normalsize{$^{7}$Kavli Institute at Cornell for Nanoscale Science, Cornell University, Ithaca, NY, USA.}\\
\normalsize{$^{8}$Leibniz-Institut f\"ur Kristallz\"uchtung, Max-Born-Str. 2 Berlin, Germany.}\\[6pt]
\\
\normalsize{$^{\ast}$To whom correspondence should be addressed; E-mail: shusain@berkeley.edu.}
}
\begin{document}
\baselineskip24pt
\maketitle

\begin{sciabstract}
Topological insulators have been explored extensively for spin–charge interconversion via magnetic interfaces, yet the true response of their spin–charge conversion, particularly in the absence of an external magnetic field, remains to be studied. Here, we report electric field control of spin-charge conversion in the topological insulator Bi$_2$Te$_3$ with the antiferromagnetic multiferroic BiFeO$_3$, employing a nonlocal spin transport device. A systematic thickness dependence of the spin transport across the interface between Bi$_2$Te$_3$ and BiFeO$_3$ reveals a signature of topological surface-state dominated spin transport in the bilayer system. The spin-charge conversion remains robust for thicknesses above 10 nm but falls rapidly with reducing thickness and vanishes at 5 nm. This is consistent with hybridization-induced emergence of a trivial insulating phase, which is supported by the coherency factor estimated from the magnetoconductance of Bi$_2$Te$_3$. These results establish that spin-momentum-locked surface states dominate interfacial spin transport in the decoupled regime. Beyond presenting efficient spin-charge interconversion at an entirely insulating magnetic interface, this work also highlights sputter-deposited Bi$_2$Te$_3$ as a high-quality and scalable platform for integrating quantum materials onto the devices. The nonlocal spin transport approach presented here provides a simple and unprecedented direct evidence of spin-charge conversion and opens an efficient and practical pathway toward designing energy-efficient spin-based devices.
\end{sciabstract}


\noindent\textbf{Introduction}\\
The rapid progress in next-generation spintronic technologies has stimulated intense interest in efficient spin-to-charge conversion (SCC) mechanisms, particularly through measurements of the inverse spin Hall effect  and the inverse Rashba-Edelstein effect in both three-dimensional (3D) and two-dimensional (2D) material systems~\cite{kato2004,saitoh2006,sinova2015}. Topological insulators (TIs), such as Bi-based chalcogenides, are especially promising in this context due to their strong spin--orbit coupling (SOC) and the spin--momentum locking inherent to their Dirac surface states~\cite{hasan2010,qi2011,zhang2009,PhysRevB.97.125414}. These unique properties enable highly efficient SCC, making TIs attractive platforms for low-power spin-based devices \cite{hoque2024room}.

Most previous studies of spin transport and SCC in TI-based heterostructures have employed metallic ferromagnets as spin sources~\cite{mellnik2014,fan2014,shiomi2014,jamali2015,kondou2016,hoque2024room,Noyan2025}. However, the presence of conductive ferromagnetic layers introduces parasitic charge current shunting and additional interfacial spin-dependent scattering processes, which can suppress the intrinsic response of the TI surface states and complicate quantitative analysis \cite{carlosPRLTI}. Moreover, metallic interfaces can partially screen, hybridize with, or otherwise perturb the topological surface states, thereby reducing their dominant contribution to spin transport. Also, a magnetic field has been extensively used as a control knob for probing the efficiency in TIs \cite{otanimaekawaTI}; switching to an electric field would, in addition to the new physics,  facilitate integration into miniaturized elements.

To overcome these limitations, insulating magnets, particularly ferroelectric antiferromagnets, provide a compelling alternative. In this work, we investigate spin transport in heterostructures composed of the topological insulator Bi$_2$Te$_3$ interfaced with the antiferromagnetic insulator BiFeO$_3$ employing non-local spin transport geometry \cite{sundararajan2025toward,cornelissen2015long,
Husain2024Non-volatileMultiferroic}. Such antiferromagnetic insulators offer additional advantages, including negligible stray fields, ultrafast spin dynamics, and robustness against external magnetic perturbations~\cite{jungwirth2016,baltz2018}. Finally, BiFeO$_3$ being a multiferroic, gives an additional degree of freedom to control the spin transport by an electric field. The Bi$_2$Te$_3$/BiFeO$_3$ heterostructure thus provides every possible ingredient as a clean and electrically insulating platform to probe interfacial spin transport mediated primarily by topological surface states.
\begin{figure*}[t]
    \centering
    \includegraphics[width=1\linewidth]{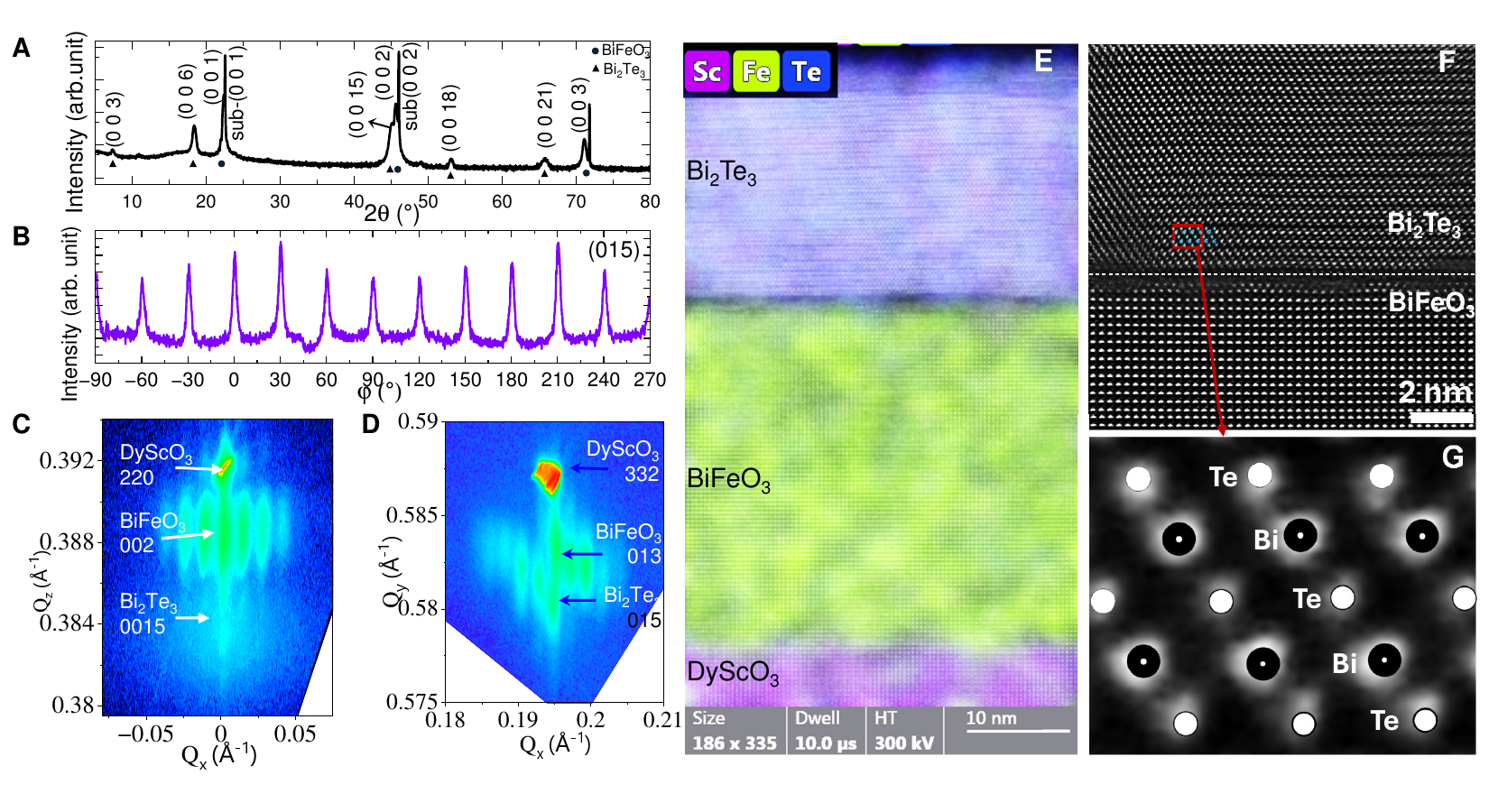}
    \caption{\textbf{Epitaxial growth of Bi$_2$Te$_3$ on BiFeO$_3$:} (\textbf{A}) X-ray diffraction line scan in $\theta-2\theta$ geometry of Bi$_2$Te$_3$/BiFeO$_3$ heterostructure. (\textbf{B}) $\varphi$ - scan of Bi$_2$Te$_3$ corresponding to the 015 reflection. Reciprocal space map of Bi$_2$Te$_3$/BiFeO$_3$ along the (\textbf{C}) symmetric 220 and (\textbf{D}) asymmetric 332 plane of DyScO$_3$ substrate. (\textbf{E}) HAADF cross-sectional TEM chemical mapping of Bi$_2$Te$_3$/BiFeO$_3$ bilayer sample showing high quality crystallinity with zoomed data in (\textbf{F}), with atomic view of Bi$_2$Te$_3$ in (\textbf{G}).}
    \label{fig:structural}
\end{figure*}

\noindent\textbf{Results}\\
We first establish the high-quality of the  Bi$_2$Te$_3$ thin films, achieved at a substrate temperature as low as 150$^\circ$C (Methods). Despite the non-equilibrium nature of sputtering, the resulting films exhibit well-defined crystallinity and thickness control suitable for investigating topological charge/spin transport. We fabricate the devices of heterostructures consisting of Bi$_2$Te$_3$ with systematically varied thicknesses deposited on $\sim$50 nm thick epitaxial BiFeO$_3$ grown on DyScO$_3$ substrates, forming an oxide/van der Waals heterostructure. This architecture enables the study of direct spin–charge conversion in the topological insulator without parasitic current shunting or screening effects in the absence of a magnetic field. Devices are patterned into a nonlocal spin-transport geometry using optical lithography, allowing purely electrical detection of spin-to-charge conversion mediated by the spin–momentum–locked states. Comprehensive structural characterization confirms the high crystalline quality and interface integrity, while systematic spin transport measurements reveal the evolution of the spin–charge conversion as a function of Bi$_2$Te$_3$ thickness.

Symmetric $\theta$/2$\theta$ scan (Fig.~\ref{fig:structural}A) reveals the Bi$_2$Te$_3$ grew (0 0 $l$) oriented on (0 0 1)BiFeO$_3$ surface. The lattice parameter $c$ was found to be 30.13 $\text{\AA}$. To identify the crystallographic ordering, $\varphi$ - scan (Fig.~\ref{fig:structural}B) was conducted along the asymmetric (0 1 5) plane for which the goniometer setting was kept at $\chi$=56.43$\degree$, 2$\theta$=27.8348$\degree$, $\omega$=9.2301$\degree$, $\omega_{offset}$=-4.6873$\degree$. $\varphi$ - scan reveals twelve distinct diffraction peaks separated by $30^\circ$, indicating the coexistence of two in-plane rotational variants (Fig.S1 and Fig.S2). Each variant preserves the intrinsic six-fold rotational symmetry of hexagonal Bi$_2$Te$_3$, resulting in an overall twelvefold symmetry. This rotational degeneracy is characteristic of graphoepitaxial growth, arising from the weak in-plane symmetry constraint imposed by the underlying rhombohedral BFO surface, and leads to the formation of $30^\circ$ rotational phase boundaries within the Bi$_2$Te$_3$ layer.

To further strengthen the epitaxial nature of Bi$_2$Te$_3$,  rocking curve along BFO (0 0 3) and Bi$_2$Te$_3$ (0 0 6) (Fig.S2) and reciprocal space mapping (RSM) maps were collected around (0 0 15) and (0 1 5) reflections (Figs.~\ref{fig:structural}C, D) of Bi$_2$Te$_3$ to quantify the lattice parameters and epitaxial registry of the heterostructure. The Bi$_2$Te$_3$ (0~1~5) reflection exhibits a broadened in-plane distribution, restricting the formation of a unique in-plane lattice parameter and consistent with a graphoepitaxial growth mode accompanied by a 30$\degree$ rotational/twin domains. The lattice parameters $a$ and $c$ are found to be 3.95~\AA, and 30.08~\AA, respectively, which is in agreement with the expected layered stacking along the $c$ axis. From the symmetric RSM acquired around the DyScO$_3$ (220) reflection, the BiFeO$_3$ (002) peak yields an out-of-plane pseudo-cubic lattice parameter of $c_{\mathrm{BFO}} \approx 3.96$~\AA, consistent with the rhombohedral BiFeO$_3$ under epitaxial constraint \cite{Meisenheimer2024DesignedAntiferromagnets}.  These reciprocal-space observations are directly corroborated by real-space imaging using high-angle annular dark-field scanning transmission electron microscopy (HAADF-STEM) (Fig.~\ref{fig:structural}E). The cross-sectional view of element-resolved atomic mapping confirms a chemically abrupt interface with minimal interdiffusion between BFO and Bi$_2$Te$_3$. Atomically resolved HAADF images reveal coherent epitaxy in the BFO layer and a graphoepitaxial Bi$_2$Te$_3$ overlayer containing sharp $30^\circ$ rotational phase boundaries (Fig. S1), consistent with the 12-fold symmetry observed in $\varphi$-scan results. Moreover, small regions of Bi-deficiency at the interface were observed, and how such interfaces could impact the spin transport will be discussed later. 


\begin{figure*}[t!]
    \centering
    \includegraphics[width=1\textwidth]{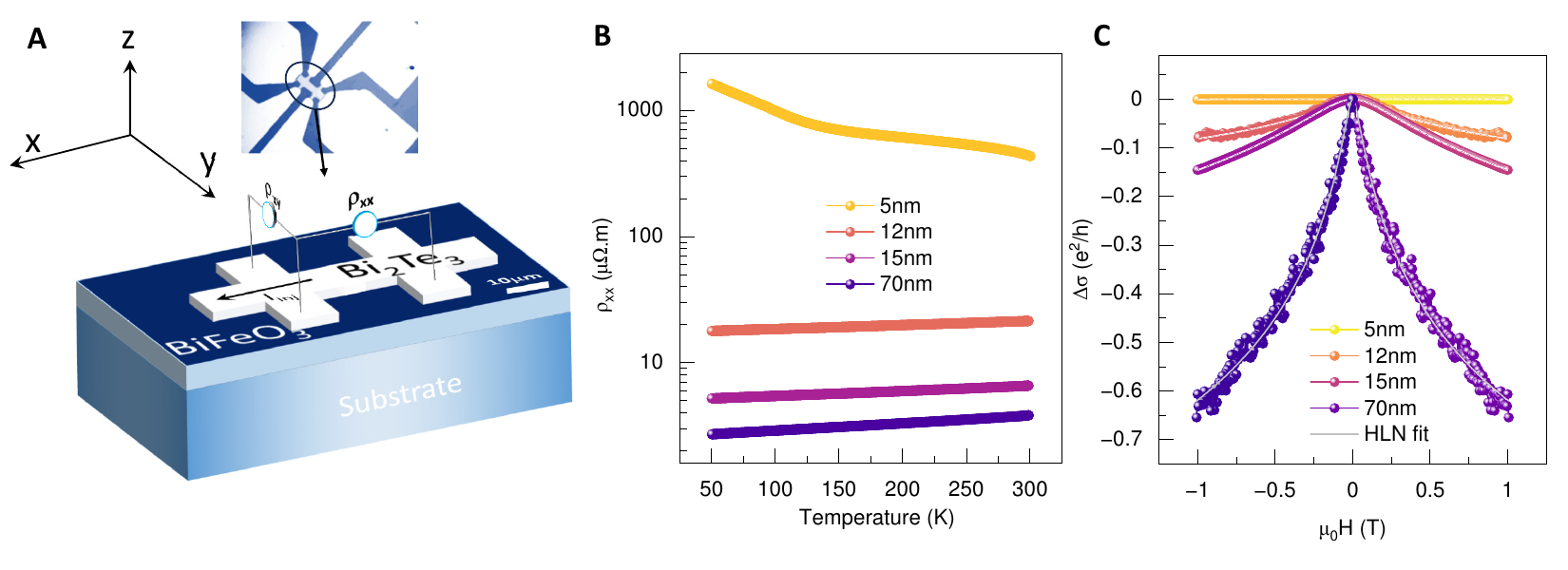}
    \caption{\textbf{Thickness‑dependent transport of Bi$_2$Te$_3$/BiFeO$_3$ heterostructures}: (\textbf{}A) Device geometry and transport measurement circuit. The inset is the optical image of the Hall-bar device. Temperature-dependent (\textbf{B}) resistivity and (\textbf{C}) magnetoconductivity measured in various Bi$_2$Te$_3$ thicknesses with fixed BiFeO$_3$. Magnetoconductivity data were recorded at a temperature of 2 K. Lines represent the fit to the Hikami–Larkin–Nagaoka (HLN) model.}
    \label{fig:transport}
\end{figure*}
The detailed transport measurements were conducted in the Hall bar geometry depicted in Fig.~\ref{fig:transport}A. The temperature-dependent longitudinal resistivity for Bi$_2$Te$_3$ thin films of thicknesses from 70 nm to 5 nm  (Fig.~\ref{fig:transport}B) shows a crossover from metallic‑like to semiconducting‑like behavior as the film becomes ultrathin. 
In this case, a resistivity upturn at low temperatures indicates a localization effect in bulk-dominated transport. If one draws a vertical line at any fixed temperature, the resistivity increases systematically as the Bi$_2$Te$_3$ thickness is reduced from 70 to 5 nm, suggesting enhanced surface and interface scattering, reduced effective carrier density, and stronger influence of disorder in thinner films. For example, for the low-temperature resistivity upturn in the 5~nm film, the resistivity data were analyzed using both variable-range hopping (VRH) and Kondo-type models (see Fig.S3). While the VRH model failed to adequately describe the data, the Kondo-type logarithmic dependence provided a reasonable fit to the low-temperature upturn. However, such behavior is not unique to Kondo scattering and can also arise from quantum correction effects, such as weak localization and electron--electron interactions, in disordered low-dimensional systems \cite{Pandey_2023}. This interpretation is further supported by the excellent agreement of the low-temperature magnetoconductance with the Hikami--Larkin--Nagaoka (HLN) weak-localization model (discussed in detail later in the text), indicating that quantum interference plays a significant role in the ultrathin film. Although a Kondo-like contribution cannot be completely excluded, the combined transport and magnetoconductance analyses suggest that disorder-induced weak localization is the dominant origin of the low-temperature resistivity upturn in the ultrathin Bi$_2$Te$_3$ film.

Further, the transverse resistance R$_{xy}$ was recorded as a function of magnetic field (Fig.~S3), which probes the density and type of charge carriers. From the linear field dependence of R$_{xy}$, we infer that transport is dominated by a single type of carrier with a negative slope, identifying electrons as the majority carriers in Bi$_2$Te$_3$/BiFeO$_3$ heterostructures. The carrier density is found to be of the order of $10^{15}$ cm$^{-2}$, slightly higher than the values typically reported for topological insulators \cite{He2011WAL,Roy2013WAL}. This enhancement can be attributed to donor‑type intrinsic defects in Bi$_2$Te$_3$ such as Te vacancies or Bi–Te antisite defects. The quality of Bi$_2$Te$_3$ remains the same or better after being deposited on BiFeO$_3$ (Fig. S3C and Fig. S3D). The Hall resistance in Bi$_2$Te$_3$ (5nm) is not measurable within the experimental accuracy due to its high resistance (Fig. S3D; however, the ferroelectric properties remain the same (discussed later).
 
\begin{figure*}[t!]
    \centering
    \includegraphics[width=1\linewidth]{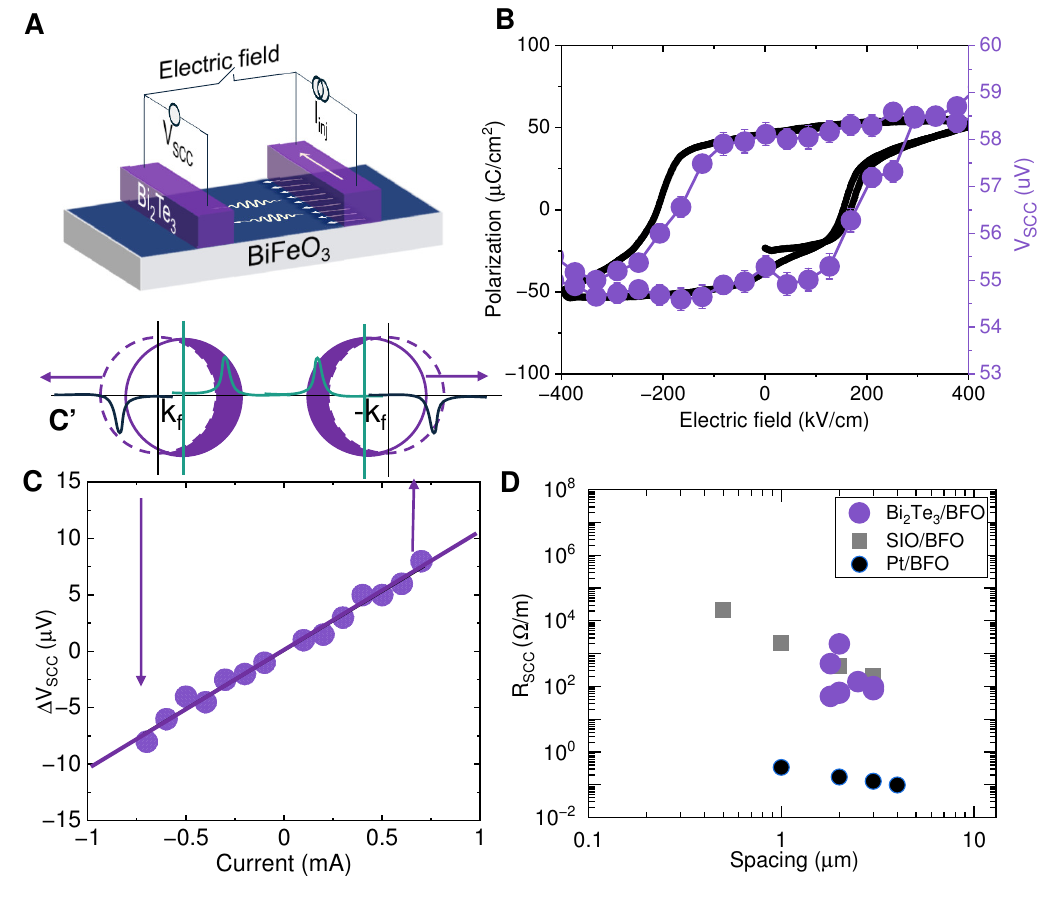}
    \caption{\textbf{Spin transport in a topological insulator:} (\textbf{A}) Non-local spin transport circuit for measuring the spin charge conversion in Bi$_2$Te$_3$ with a constant current supply with a varying electric field. (\textbf{B}) Voltage is measured as a function of the electric field across the electrode. The data in black is the ferroelectric polarization from the ferroelectric BFO, and the data in green is the voltage due to spin-charge conversion ($V_{SCC}$) measured in Bi$_2$Te$_3$ as a function of electric field. (\textbf{C}) Differential voltage $\Delta$V$_{SCC}$ measured as a function of supply current.  Drawing in the top of the plot with arrows shows the shifting the Fermi surface under the electric current supply in opposite directions. (\textbf{D}) Comparison of the SCC resistance $R_{SCC}=\frac{V_{SCC}}{I_{inj}\times L}$ (L being the length of the electrode) between the conventional materials such as Pt and SrIrO$_3$\cite{huang2024manipulating}.}
    \label{fig:nonlocal}
\end{figure*}

The low‑field magnetoconductivity $\Delta$$\sigma$($e^{2}$/h) measured at temperature 2K for range of Bi$_2$Te$_3$ thicknesses (Fig.~\ref{fig:transport}C). The comparison of magnetoresistance data measured on the two thicknesses is represented in Fig.~S4. The observed cusp-like behavior around zero transverse magnetic field exhibits the presence of weak antilocalization (WAL) in a topological system with strong spin–orbit coupling. To quantify this quantum‑interference correction, we analyze the magnetoconductivity using the Hikami–Larkin–Nagaoka (HLN) model: \cite{Hikami1980SpinOrbit}, $\Delta$$\sigma$ = -$\alpha$($e^{2}$/h) $ [ (h/{8\pi e H L_{\phi}^2}) - \psi(1/2 + h/{8\pi e H L_{\phi}^2})) ] $ where $\alpha$ is a dimensionless prefactor that indicates the number of coherent conduction channels predicted to be $\frac{1}{2\pi}$ for each 2D channel, $\psi$ is digamma function and $L_\phi$ is phase coherence length. The pre-factor $\alpha$ is estimated for different Bi$_2$Te$_3$ thicknesses. For 70nm, the $\alpha$ is -0.27, suggesting weak anti-localization dominated by a single 2D channel ($\frac{1}{\pi}$) coexisting with a finite bulk background (c.f. large carrier density). The WAL cusp becomes less pronounced for a 15 nm thick Bi$_2$Te$_3$ sample as the fitted prefactor decreases to $\alpha$ = -0.17. This indicates that the top and bottom channels ($\frac{1}{2\pi}$) are both contributing to the charge transport \cite{chaeong2011}. On the other hand, for the thinnest films (5nm), the magnetoconductivity is nearly flat within the experimental resolution, and the HLN fit yields a value that is nearly zero. The low value of $\alpha$ suggests that WAL is mostly quenched, pointing to a regime of strong localization or gap opening where coherent back-scattering corrections no longer produce a pronounced interference cusp. 
The reduced values of the WAL prefactor should also be considered in light of the unavoidable bulk contribution in Bi$_2$Te$_3$. Owing to intrinsic Te vacancies and Bi/Te antisite defects, Bi$_2$Te$_3$ typically exhibits finite bulk conductivity even in high-quality thin films. Therefore, the measured magnetoconductivity reflects parallel transport through both topological surface states and the conducting bulk, and the extracted HLN prefactor represents an effective contribution from multiple transport channels rather than isolated surface states. This surface--bulk interplay has been extensively reported in Bi$_2$Te$_3$ and is regarded as an intrinsic characteristic of the material, making it difficult to achieve purely surface-state transport without compensation doping or electrostatic gating \cite{PhysRevLett.109.066803,hoefer2014intrinsic,PhysRevB.81.241301,zhang2010crossover}.
These observations suggest that Bi$_2$Te$_3$/BiFeO$_3$ interfaces preserve the key transport features characteristic of a topological insulator, such as WAL and thickness-dependent surface–bulk interplay \cite{PhysRevB.87.035122,chaeong2011,Pandey_2023}.

Armed with the crystalline quality and charge transport behavior of the heterostructure, non-local spin transport measurements were conducted in lithographically patterned devices (Methods), as shown schematically in Fig~\ref{fig:nonlocal}A. In these devices, the injected DC $I_{inj}$, generates a polarized spin current (through spin-charge interconversion) at the interface due to spin-orbit coupling arising from the Berry curvature of the surface states and bulk, leading to a non-equilibrium chemical potential at the interface. Note that the bulk contribution cannot be avoided in Bi$_2$Te$_3$ crystal due to its Fermi level lying in the conduction band \cite{hoefer2014intrinsic,chen2009experimental,Noyan2025}, which has been observed in other topological insulators as well \cite{kondou2016fermi}. When these polarized spins transfer their angular momentum to the antiferromagnetic spin of the BiFeO$_3$, the spin gradient excites magnons in BiFeO$_3$, creating a magnon population gradient that drives them toward the detector. The resulting spin accumulation at the detector interface produces a voltage through spin-charge conversion via the inverse Edelstein effect \cite{carlosPRLTI}. We represent the measured voltage as $V_{SCC}$, which is the net effect from surface and bulk. We record the $V_{SCC}$ as a function of external electric field applied across the two Bi$_2$Te$_3$ electrodes to change the ferroelectric polarization in BiFeO$_3$ shown in Fig.~\ref{fig:structural}B. The data in black is the ferroelectric polarization, and the green hysteresis corresponds to the non-local $V_{SCC}$. The $V_{SCC}$ is recorded in the remanent state of the polarization measured after an electric field pulse; the data was recorded for about 10 seconds at constant DC, giving a data point with standard deviation as the error plotted. Therefore, the data are free of artifacts due to the electric field. The $V_{SCC}$  exactly follows the ferroelectric polarization, which is indicative of the electric-field-controlled spin transport. This is also consistent with earlier works on electric field-controlled antiferromagnetism reported with the conventional material systems \cite{Meisenheimer2024DesignedAntiferromagnets, Husain2024Non-volatileMultiferroic, huang2024manipulating,Chai2024VoltageLogic-in-memory}. We would like to emphasize that the measured voltage $V_{SCC}$ is close to what has been reported for SrIrO$_3$ using the same method with the SCC efficiency of $\sim$0.3-0.4 \cite{huang2024manipulating}; therefore, we can correlate the similar magnitude of spin Hall angle in Bi$_2$Te$_3$, which is close to the theoretical value estimated from density functional theory calculations \cite{PhysRevMaterials.4.114202}. Since we are measuring the SCC through a direct approach without a magnetic field, therefore, artifact related to the mathematical calculations and magnetic field-based analysis are avoided. Here we suggest that by tuning the band structure of the topological insulators such as Bi$_2$Se$_3$ or Bi$_x$Sb$_{1-x}$, etc (as per reported works\cite{Noyan2025}), $V_{SCC}$ can be amplified for device applications, for example, to reach $V_{SCC}$=100 mV for MESO device \cite{Manipatruni2019MagnetoelectricCharge}.

A detailed current polarity dependence was executed to understand the origin of the $V_{SCC}$. If the $V_{SCC}$ were to have an origin from the thermal gradient (Seebeck effect) or thermal magnons, the hysteresis would not reverse the polarity \cite{Husain2024Non-volatileMultiferroic}. However, it does if the charge-spin conversion is right at the source and spin-charge conversion is at the detector \cite{Husain2025ColossalAntiferromagnet}. To further substantiate the thermal origin of the magnon transport, second-harmonic measurements were carried out using a lock-in detection technique, with the voltage signal measured at $2\omega$. The corresponding data is presented in Fig.~S10 and is consistent with the presence of thermally generated magnons in the BFO/Bi$_2$Te$_3$ heterostructure. The resistance of the magnonic medium BiFeO$_3$ is of the order of G$\Omega$, which protects against any charge current leakage to the detector. Therefore, the current-driven $V_{SCC}$ polarity reversal has only the source right at the high spin orbit coupling from the interface/bulk of the topological insulator. The differential voltage $\Delta V_{SCC}$ is the peak-to-peak voltage recorded from the hysteresis plots for a range of current values shown in Fig.~\ref{fig:nonlocal}C.  $\Delta V_{SCC}$ reverses its sign when it goes from negative - to positive + (see Fig.~5).

This sign reversal has been typically discussed theoretically in topological insulators in terms of the current-induced shift of the Fermi surface of the helical Dirac surface state, as schematically depicted in Fig.~\ref{fig:nonlocal}C'. Within the semiclassical Boltzmann picture (Methods), an applied electric field (or electrical current in our case) displaces the Fermi contour in momentum space by $\Delta \mathbf{k} \propto \mathbf{E}$, producing an imbalance between states at $+\mathbf{k}$ and $-\mathbf{k}$. Here, E is the electrical current supplied to the Bi$_2$Te$_3$ wire. Because of spin–momentum locking, where the spin expectation value satisfies $\langle \mathbf{s}(\mathbf{k}) \rangle \propto \hat{\mathbf{z}}\times \hat{\mathbf{k}}$, this asymmetric occupation generates a non-equilibrium spin density $\mathbf{S} \propto \hat{\mathbf{z}}\times \mathbf{E}$ (Edelstein effect) \cite{hasan2010} \cite{qi2011}. See details in Methods. Reversing the current direction reverses the Fermi-surface shift, thereby flipping the direction of the induced spin polarization and consequently the sign of the spin–charge conversion voltage $V_{SCC}$. In this surface-dominated scenario, the polarity of $V_{SCC}$ is therefore directly determined by the helicity of the Dirac cone.

On the other hand, a sign reversal of $V_{SCC}$ can also arise if the bulk spin–orbit coupling provides a significant contribution to the conversion process. In that case, the inverse spin Hall effect in the bulk generates an electric field $\mathbf{E}{\mathrm{_{ISHE}}} \propto \theta{\mathrm{_{SH}}} (\mathbf{J_s} \times \hat{\boldsymbol{\sigma}})$, where $\theta{\mathrm{_{SH}}}$ is the bulk spin Hall angle. The measured voltage then reflects the combined contributions from both the surface inverse Edelstein and bulk inverse spin Hall mechanisms, $V_{SCC} \propto (\eta_{\mathrm{surf}} + \eta_{\mathrm{bulk}}) I_{inj}$. Here $\eta$ is the material parameter. If these two coefficients have opposite signs due to their distinct microscopic origins, which can drive the sign of the total conversion efficiency, leading to a sign change in $V_{SCC}$. Here we are reversing the direction of the injected current $I_{inj}$, which produces the opposite voltage (Methods \ref{B}).

For a quantitative comparison, the data obtained from various devices and samples are plotted alongside those of reported conventional heavy metal Pt and the oxide spin–orbit metal SrIrO$_3$ \cite{huang2024manipulating}. This suggests that the large spin–charge conversion voltage magnitude in Bi$_2$Te$_3$ is orders of magnitude larger than that of Pt. Notably, our measurements are insensitive to the resistance of the metallic layer (as discussed below), since the detected signal corresponds to the voltage difference between the two electrically controlled antiferromagnetic states of BiFeO$_3$. The magnitude of $\Delta V_{SCC}$ is comparable to, or even exceeds, that of existing materials, including state-of-the-art all-epitaxial Bi$_2$FeO$_3$/SrIrO$_3$ heterostructures \cite{huang2024manipulating,Meisenheimer2024DesignedAntiferromagnets}.
\begin{figure}[t]
    \centering
    \includegraphics[width=0.45\textwidth]{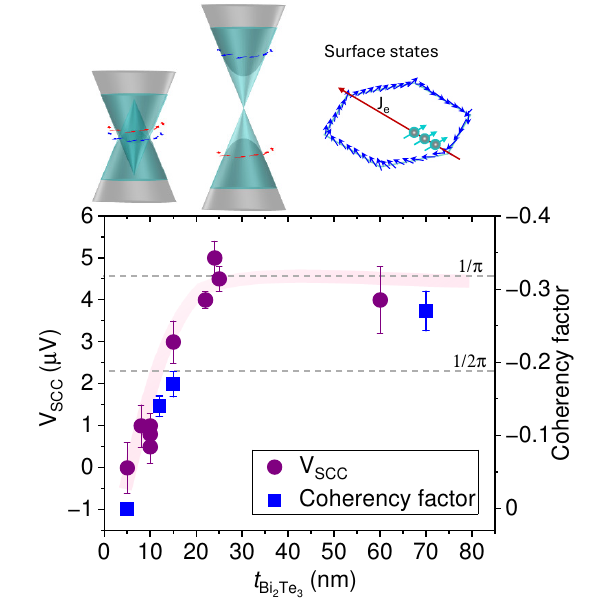}
    \caption{Spin charge conversion and the influence of topological behavior: Differential spin charge conversion ($\Delta V_{SCC}$) measured between the two (+$P$ and - $P$) polarization states as a function of Bi$_2$Te$_3$ thickness. The data in red squares represent the coherency factor estimated from the weak anti-localization for several thicknesses. A shaded line is drawn to guide the eye. Dotted lines represent the case where a single and double 2D channel from the surface states governs the transport.}
    \label{fig:analysis}
\end{figure}

To understand the origin of the signal, we extended our measurements for a range of Bi$_2$Te$_3$ thicknesses since the thickness controlled the properties of the topological insulators \cite{chen2009experimental}. The $V_{SCC}$ for various thicknesses of Bi$_2$Te$_3$ is plotted in Fig.~\ref{fig:analysis}. The $V_{SCC}$ remains nearly constant for Bi$_2$Te$_3$ thicknesses down to $\sim$20 nm, indicative of being dominated by the surface, followed by a gradual reduction at 15 nm and a pronounced suppression at 10 nm. Below $\sim$8 nm, the voltage output becomes negligible. Though the resistance of the lowest thickness is the largest (Fig.~\ref{fig:transport}A), it does not influence our spin transport. The ferroelectric hysteresis remains the same at lower Bi$_2$Te$_3$ thicknesses (Fig.~S4), which reflects the robust ferroelectric and, consequently, the antiferromagnetic state in BiFeO$_3$. Thus, the thickness dependence in the SCC reflects the evolution of the electronic transport channels in Bi$_2$Te$_3$. The extracted coherency factor, plotted as a function of thickness, shows excellent agreement with the spin-transport data, indicating that the efficiency of spin–charge conversion is dominated by the topological surface states. The dotted reference lines marked as $1/\pi$ and $1/2\pi$ correspond to dual and single two-dimensional (2D) conduction channels, respectively, consistent with weak anti-localization analysis. In thicker films, transport is dominated by the two decoupled top and bottom surface states, enabling robust inverse Edelstein conversion. As the thickness decreases, hybridization between the two surfaces and increased bulk scattering reduce the effective spin-momentum locking contribution. In the ultrathin limit, although the resistivity increases due to enhanced disorder and possible gap opening from surface-state hybridization, the suppression of the topological nature of Bi$_2$Te$_3$ eliminates efficient spin–charge interconversion. The conventional thickness dependence observed here suggests that spin transport is primarily governed by bulk conduction, as surface-state contributions are expected to be independent of thickness.

In conclusion, we demonstrate direct evidence of spin–charge interconversion in the topological insulator Bi$_2$Te$_3$, electrically controlled via BiFeO$_3$, which serves as an antiferromagnetic spin-transport medium. For the first time, spin–charge conversion is modulated by an electric field through antiferromagnetic magnons in a topological insulator, providing deterministic control without relying on charge-current switching or magnetic field. The pronounced thickness dependence of the spin transport suggests signatures of topological surface states, further supported by weak antilocalization behavior consistent with coherent two-dimensional conduction. The magnitude of the measured voltage output is orders of magnitude larger than that of conventional heavy metals such as Pt, providing compelling evidence of the superiority of topological insulators for efficient spin–charge conversion. These findings highlight the technological potential of topological insulators for low-power, electrically tunable spin-based devices.

\vspace{1 cm}


\noindent\textbf{Materials and Methods}
\begin{flushleft}
\textbf{Thin film deposition} 
\end{flushleft}

\noindent\textbf{Bi$_2$Te$_3$}\\
Bi$_2$Te$_3$ thin films were grown using RF magnetron sputtering from a single high-purity (99.99$\%$) target. Prior to deposition, the substrates were heated to growth temperature ($150^{\circ}\mathrm{C}$). The Bi$_2$Te$_3$ layer was then deposited on BiFeO$_3$ in a ultra high-vacuum chamber (base pressure of $\sim 1-3 \times 10^{-8}$ Torr) with Ar flow rate of 25 sccm, and the RF power was set to 10 Watt. After growth, the samples were cooled down to room temperature in a vacuum, and then a thin Te capping layer ($\sim$1 nm) was deposited at room temperature to protect the Bi$_2$Te$_3$ surface from degradation.\\
\noindent\textbf{BiFeO$_3$}\\
Epitaxial BiFeO$_3$ (BFO) thin films were grown on DyScO$_{3}$(220) substrates by pulsed laser deposition (PLD) and Molecular beam epitaxy (MBE). For deposition by PLD, we have used a commercially available BiFeO$_3$ ceramic target. The deposition was performed at a substrate temperature of $720^{\circ}\mathrm{C}$ under an oxygen pressure of $150\,\mathrm{mTorr}$, the laser energy density was maintained $1.2\,\mathrm{J/cm}^2$ with a repetition rate of $10\,\mathrm{Hz}$. MBE films were grown by reactive MBE in a VEECO GEN10 system using a mixture of 80 \%  ozone (distilled) and 20 \% oxygen. Elemental sources of bismuth and iron were used at fluxes of \SI{1.5e14}{} and  \SI{2e13}{atoms \per\centi\meter\squared\second} respectively. All films were grown at a substrate temperature of \SI{675}{\celsius} and chamber pressure of 5 x 10$^{-6}$ Torr. The samples thickness and qualities were maintained irrespective of the deposition process.  \\
\noindent\textbf{Structural Characterization}\\
The crystal structures of both Bi$_2$Te$_3$ and BiFeO$_3$ were determined through X-ray diffraction, utilizing a high-resolution X-ray diffractometer (PANalytical, X’Pert Pro). The symmetric line scan ($\theta$–2$\theta$) employed a fixed-incident-optics slit set at 1/2$^\circ$, while the reciprocal space mapping (RSM) involved an asymmetric 2D scan with a slit of 1/4$^\circ$. The X-ray source was Cu K$\alpha$ transition (wavelength: 1.5401 \AA), and detection employed a PIXcel$^{3D}$-Medipix$^3$ detector with a fixed receiving slit of 0.275 mm.\\
\noindent\textbf{Device Fabrication}\\
Post film depositions, non-local devices were patterned using photolithography using a Maskless Aligner MLA-150. Following exposure, the resist underwent wet-etching using MEGAPOSIT MF-26A photoresist developer for 18 seconds. Non-local device (Length, 100$\mu$m | width, 1.5$\mu$m with spacing 1-3$\mu$m) and Hall bar (60$\mu$m$\times$10$\mu$m) devices made of Bi$_2$Te$_3$ were etched down via an Ar ion mill down to BiFeO$_3$ surface. The milling was carried out at an angle of 30$\degree$ with a beam voltage of 300 V, and a beam current of 75 mA. During the milling process, the heterostructures were placed on a sample stage that spun at a constant speed of 15 rpm to achieve uniform milling. The temperature was maintained at 15º C to avoid sample heating during the ion-milling process. The etch rate was first calibrated using an atomic force microscope on a single layer Bi$_2$Te$_3$ film and confirmed by ferroelectric measurements in fabricated devices. Pt metal was used as a contact prepared by a lift-off process. The lithography was performed at the Marvell Nanofabrication facility at Berkeley.\\
\noindent\textbf{Hall Effect and Magnetoconductance}\\
Electrical transport measurements were performed using a standard six-terminal Hall bar geometry fabricated on as-grown thin films. Both the longitudinal and transverse (Hall) resistance were measured using a CRYOGENIC and Quantum design magnet system in a transverse magnetic field. A Keithley 2400 source meter was used to supply a constant DC along the device channels, while the corresponding longitudinal and transverse voltage drops were recorded using a Keithley 2182A nanovoltmeter.\\
\noindent\textbf{Spin Transport Measurements}\\
Spin transport measurements were conducted employing four-terminal non-local devices, wherein two terminals were dedicated to source current injection (using 6221 Keithley) for charge to spin current conversion and the remaining two served as output terminals to measure the spin to charge conversion. One source terminal and one detection terminal were also used to apply an electric field for ferroelectric polarization control, while current and voltage leads were turned off. The measurements were conducted in the remanent state means no electric field was applied during the measurement, thus eliminating any interference during the magnon transport. Also, no magnetic field was applied in these measurements. The entire experimental setup and procedures were controlled using an in-house developed Python code and a Keithley 7001 switch box, maximizing repeatability.\\

\noindent\textbf{Ferroelectric Measurements}\\
To assess the ferroelectric nature of the epitaxial films, we fabricated both in-plane capacitor devices on Bi$_2$Te$_3$/BiFeO$_3$. The in-plane devices' electrodes were patterned by photolithography. Polarization–electric field (P–E) hysteresis loops were recorded using a Precision Multiferroic system (Radiant Technologies).\\

\noindent\MakeUppercase{\textbf{Microscopic Origin of the Sign Reversal in Spin--Charge Conversion}}
\subsection{Fermi-Surface Shift and Nonequilibrium Spin Polarization}
From the semiclassical Boltzmann formalism under the relaxation-time approximation, when an in-plane electric field $\mathbf{E}$ is applied, the distribution function acquires a first-order correction ,
\begin{equation}
f(\mathbf{k}) = f_0(\varepsilon_{\mathbf{k}})
- e \tau (\mathbf{E}\cdot \mathbf{v}_{\mathbf{k}})
\left(-\frac{\partial f_0}{\partial \varepsilon}\right),
\label{eq:boltzmann}
\end{equation}
where $\tau$ is the momentum relaxation time and 
$\mathbf{v}_{\mathbf{k}} = \frac{1}{\hbar}\nabla_{\mathbf{k}}\varepsilon_{\mathbf{k}}$ 
is the velocity. This is equivalent to a rigid shift of the Fermi contour,
\begin{equation}
\Delta \mathbf{k} = -\frac{e\tau}{\hbar}\mathbf{E}.
\label{eq:kshift}
\end{equation}

For a topological surface state, the low-energy Hamiltonian is
\begin{equation}
H = \hbar v_F (\hat{\mathbf{z}}\times\boldsymbol{\sigma})\cdot\mathbf{k},
\label{eq:dirac}
\end{equation}
which describes a helical Dirac cone with spin–momentum locking 
\cite{hasan2010,qi2011}. The expectation value of spin is represented as, $\langle \mathbf{s}(\mathbf{k}) \rangle 
= \frac{\hbar}{2}(\hat{\mathbf{z}}\times \hat{\mathbf{k}}).$

The nonequilibrium spin density induced by the shifted distribution is
\begin{equation}
\mathbf{S}
= \sum_{\mathbf{k}} 
\langle \mathbf{s}(\mathbf{k}) \rangle
\delta f(\mathbf{k}),
\label{eq:spindensity}
\end{equation}
where $\delta f = f - f_0$. Substituting Eq.~(\ref{eq:boltzmann}) and evaluating the angular integral over the Fermi circle yields
\begin{equation}
\mathbf{S}
= \frac{e \tau D(E_F) v_F^2}{2} (\hat{\mathbf{z}}\times \mathbf{E})
\label{eq:edelstein}
\end{equation}
Here $D(E_F)$ is the density of states at the Fermi energy.
This represents the Edelstein effect (inverse spin galvanic effect) 
\cite{Edelstein1990,Shen2014}.
Here $E$ is treated as an effective electric field through the injected current $J_c$ at the interface. Therefore,
\begin{equation}
\mathbf{J_c} \rightarrow -\mathbf{J_c}
\quad \Longrightarrow \quad
\mathbf{S} \rightarrow -\mathbf{S},
\end{equation}
because the Fermi-surface shift $\Delta \mathbf{k}$ reverses direction.
This provides a phenomenological description of the sign reversal:
reversing current reverses the direction of the induced spin polarization.

\subsection{Conversion from Spin Accumulation to Measured Voltage}\label{B}

If spin accumulation produces a transverse electric field through 
inverse Edelstein or inverse spin Hall processes, the spin–charge 
conversion field may be written phenomenologically as,
\begin{equation}
\mathbf{E}_{\mathrm{SCC}} 
= \lambda_{\mathrm{SCC}} 
(\hat{\mathbf{z}}\times \mathbf{S}),
\label{eq:sccfield}
\end{equation}
where $\lambda_{\mathrm{SCC}}$ is a conversion coefficient. From Eq.~(\ref{eq:edelstein}) and (\ref{eq:sccfield}), $\mathbf{E}_{\mathrm{SCC}}
\propto 
\hat{\mathbf{z}}\times 
(\hat{\mathbf{z}}\times \mathbf{E})$.
Using the vector identity 
$\mathbf{a}\times(\mathbf{b}\times\mathbf{c})
= \mathbf{b}(\mathbf{a}\cdot\mathbf{c})
- \mathbf{c}(\mathbf{a}\cdot\mathbf{b})$, $\hat{\mathbf{z}}\times 
(\hat{\mathbf{z}}\times \mathbf{E})
= -\mathbf{E},$ for in-plane $\mathbf{E}$.
Thus the voltage over length $L$ along direction $\hat{\ell}$ can be measured as
\begin{equation}
V_{SCC} 
= L \, \hat{\ell}\cdot \mathbf{E}_{\mathrm{SCC}}
= \eta I,
\label{eq:linearI}
\end{equation}
where $\eta$ can be defined as the material parameter. This can explain
\begin{equation}
V_{SCC}(-I) = -V_{SCC}(I).
\end{equation}
Therefore, the sign reversal of $V_{SCC}$ directly reflects 
the reversal of the Fermi-surface shift and induced spin polarization.

\subsection{Competition Between Surface and Bulk Contributions}
In realistic samples, both surface inverse Edelstein effect (IEE) and bulk inverse spin Hall effect (ISHE) contribute, $V_{SCC} 
= (\eta_{\mathrm{surf}} 
+ \eta_{\mathrm{bulk}}) I$.
The bulk ISHE contribution can be written as, $\mathbf{E}_{\mathrm{ISHE}} 
= \theta_{\mathrm{SH}} \rho 
(\mathbf{J}_s \times \hat{\boldsymbol{\sigma}})$, where $\theta_{\mathrm{SH}}$ is the spin Hall angle and 
$\rho$ the resistivity.
If $\eta_{\mathrm{surf}}$ and $\eta_{\mathrm{bulk}}$ have opposite signs,
a sign reversal occurs when
$\eta_{\mathrm{surf}} 
+ \eta_{\mathrm{bulk}}$ is deviated from 0 to either positive or negative.

Such sign changes can arise from Fermi-level tuning, which modifies the relative weights of 
surface and bulk transport channels. In our case, we conduct the experiment under the current $I$ reversal, which directly helps us identify the source of spin-charge conversion, as discussed in Equation.\ref{eq:linearI}.
Hence, the experimentally observed sign reversal may originate either 
from the intrinsic helicity of the surface Dirac cone or from a crossover between competing spin–orbit coupling mechanisms from surface and bulk.

\bibliography{apssamp}

\bibliographystyle{Science}

\noindent\textbf{Acknowledgments}\\
We would like to express our sincere gratitude to Professor Nitin Samarth for the stimulating discussions and constructive feedback. This work was primarily supported by the U.S. Department of Energy, Office of Science, Office of Basic Energy Sciences, Materials Sciences and Engineering Division under Contract No. DE-AC02-05CH11231 within the Quantum Materials Program (No. KC2202). S.H. and R.R. acknowledge support from DARPA Fast and Curious Program under Agreement No. HR0011269E210. S.H. D.G.S., M.R., and R.R. acknowledge support from the Army Research Office under the ETHOS MURI via cooperative agreement W911NF-21-2-0162 and the Army Research Laboratory under Cooperative Agreement Number W911NF-24-2-0100. The views and conclusions contained in this document are those of the authors and should not be interpreted as representing the official policies, either expressed or implied, of the Army Research Laboratory or the U.S. Government. The U.S. Government is authorized to reproduce and distribute reprints for Government purposes, notwithstanding any copyright notation herein.\\

\noindent\textbf{Author contributions}\\
Conceptualization: SH, RR, 
Methodology: SH, YK, PG, XL, RM, RC, MF, AO, MR, NR, 
Investigation: SH, YK, 
Visualization: SH, RR, DGS, AL, RJB, JGA, 
Supervision: SH, RR, 
Writing—original draft: SH and YK,
Writing—review $\&$ editing: SH, RR.\\

\noindent\textbf{Competing interests:}\\
All other authors declare they have no competing interests.\\

\noindent\textbf{Data and materials availability:}\\
All data are available in the main text or the supplementary materials.

\noindent\textbf{Supplementary Materials}
Supplementary Text
Figs. S1 to S12
References
\end{document}